\documentclass[11pt,a4paper]{article}
\pdfoutput=1

\usepackage{jheppub}
\usepackage[utf8]{inputenc}
\usepackage{graphicx}
\usepackage{epsfig}
\usepackage{subfigure}
\usepackage{color}
\usepackage{comment}
\usepackage{slashed}

\def\beq{\begin{equation}}
\def\eeq{\end{equation}}
\def\baq{\begin{eqnarray}}
\def\eaq{\end{eqnarray}}

\newcommand{\be}{\begin{equation}} 
\newcommand{\ee}{\end{equation}}
\newcommand{\bea}{\begin{eqnarray}} 
\newcommand{\eea}{\end{eqnarray}}

\newcommand{\bmp}{\noindent\begin{minipage}{16cm}}
\newcommand{\emp}{\end{minipage}\vskip 7mm} 
\def\lsim{\mathrel{\raise.3ex\hbox{$<$\kern-.75em\lower1ex\hbox{$\sim$}}}}
\def\gsim{\mathrel{\raise.3ex\hbox{$>$\kern-.75em\lower1ex\hbox{$\sim$}}}}

\newcommand{\intron}[1]{}

\subheader{\small{Preprint: HIP-2017-05/TH}}

\title{WIMP miracle of the second kind}

\author[a,b]{Matti Heikinheimo}
\author[c]{, Tommi Tenkanen}
\author[a,b]{and Kimmo Tuominen}

\affiliation[a]{Department of Physics, University of Helsinki \\
                      P.O.~Box 64, FI-00014, Helsinki, Finland}
\affiliation[b]{Helsinki Institute of Physics, \\
                      P.O.~Box 64, FI-00014, Helsinki, Finland}
\affiliation[c]{Astronomy Unit, Queen Mary University of London, \\
                      Mile End Road, London, E1 4NS, U.K.}

\emailAdd{matti.heikinheimo@helsinki.fi}
\emailAdd{t.tenkanen@qmul.ac.uk}
\emailAdd{kimmo.i.tuominen@helsinki.fi}

\abstract{We study dark matter production in scenarios where a scale invariant
hidden sector interacts with the Standard Model degrees of freedom via a Higgs
portal $\lambda \Phi^\dagger\Phi s^2$. If the hidden sector is very weakly
coupled to the SM but exhibits strong interactions within its own particle
species, the dark matter abundance may arise as a result of a dark freeze-out
occurring in the hidden sector. Because of scale invariance, the free
parameters in the hidden sector are determined and the dark matter candidate
exhibits a 'WIMP miracle of the second kind'. Demonstrating the predictive
power of scale invariance, we carry out thorough analysis of dark matter
production in several benchmark scenarios where the hidden sector contains
either a scalar, fermion (sterile neutrino), or vector dark matter, and discuss the observational consequences of these scenarios.
}


\begin{document}
\maketitle

%
\section{Introduction}

The existence of a significant
dark matter (DM) component in the energy density of the universe seems
indisputable \cite{Ade:2015xua}. Furthermore, the gravitational effects DM has
on cosmic structure formation provide for an efficient probe on the origin and
properties of DM. The classic examples, possibly signaling the existence of DM
self-interactions, include observations of flat cores of galactic DM
halos~\cite{Moore:1994yx,Flores:1994gz}, the so-called missing satellites
\cite{Klypin:1999uc} and the too-big-to-fail problems
\cite{BoylanKolchin:2011de}.
Self-interacting DM has been claimed to account also for the spatial offset
between the DM halo and the stars in a galaxy in the Abell 3827
cluster~\cite{Massey:2015dkw,Kahlhoefer:2015vua,Heikinheimo:2015kra,Campbell:2015fra,Sepp:2016tfs}. Presence of
such non-zero DM self-interactions is also favored by the non-observation of
isocurvature perturbations in the
CMB~\cite{Kainulainen:2016vzv,Heikinheimo:2016yds}, resulting from an
interesting interplay between inflationary dynamics at high scales and DM
production around the electroweak scale \cite{Enqvist:2014zqa, Nurmi:2015ema}.

Recently, there has been a lot of interest in models which have no
explicit scales present at the tree level.
These models have been applied to address the questions concerning the
dynamical generation of electroweak scale or the scale of dark matter \cite{Bardeen:1995kv,Meissner:2006zh,Chang:2007ki,Hambye:2007vf,Foot:2007iy,AlexanderNunneley:2010nw,Hur:2011sv, Englert:2013gz,Farina:2013mla, Heikinheimo:2013fta,Gabrielli:2013hma, Carone:2013wla, Hambye:2013sna,Foot:2013hna, Davoudiasl:2014pya, DiChiara:2015bua, Ahriche:2015loa,Karam:2015jta, Marzola:2016xgb,Kannike:2016wuy, Karam:2016rsz, Brivio:2017dfq,Marzola:2017jzl}. For example, theories where the Higgs boson appears as
a composite particle explain the origin of the electroweak scale
dynamically with essentially no fine tuning of the parameters in the
underlying gauge theory.
Scale invariant hidden sectors are natural in isolation and, if the coupling between the Standard Model (SM) and a hidden sector is small, also protected from any naturalness problems originating from the SM sector \cite{Foot:2013hna}.

If the coupling between the two sectors is very small, the DM particles never thermalize
with the SM particles and the DM abundance has to be produced non-thermally by
the so-called freeze-in mechanism instead of the usual freeze-out picture. This was originally studied in~~\cite{McDonald:2001vt} in the context of a weakly interacting
scale invariant hidden sector coupled to the SM sector via Higgs
portal $\lambda_{\rm hs}\Phi^\dagger\Phi s^2$.
The freeze-in production of dark matter requires very
small values of the portal coupling, $\lambda_{\rm hs}\simeq
10^{-10}$, and the corresponding DM particles have been called
FIMP's (Feebly Interacting Massive Particles). Recently, they have been studied in a number of different contexts, see e.g. ~\cite{McDonald:2001vt,Hall:2009bx, Yaguna:2011qn, Chu:2011be,
Klasen:2013ypa, Blennow:2013jba,Merle:2013wta,
Enqvist:2014zqa,Adulpravitchai:2014xna,Merle:2014xpa,Elahi:2014fsa,Kang:2014cia,Kang:2015aqa,Nurmi:2015ema,Merle:2015oja,Bernal:2015ova,Bernal:2015xba,Shakya:2015xnx,McDonald:2015ljz,
Gabrielli:2015hua, Kainulainen:2016vzv,Heikinheimo:2016yds,Tenkanen:2016twd,Konig:2016dzg,Bernal:2017mqb}.

The scale invariant models are highly constrained. For instance,
if DM is a singlet scalar
produced by Higgs decays, the only parameter relevant for DM
production, $\lambda_{\rm hs}$, becomes fixed as a function of the DM mass by the requirement that
the singlet $s$ comprises all of the observed DM abundance,
$\Omega_{\rm s}h^2\simeq 0.12$.
In~\cite{McDonald:2001vt,Kang:2014cia,Kang:2015aqa} it was noted that
in some parts of the parameter space these models not only yield the
correct DM abundance but can also accommodate relatively large DM
self-interactions, $\sigma/m_{\rm s}\simeq\mathcal{O}(0.1) {\rm cm}^2/{\rm g}$,
favored by observations of small scale structure
formation~\cite{Moore:1994yx,Flores:1994gz}.
Consequently, also $\lambda_{\rm s}$ is fixed.
Due to the absense of free parameters, the scenario has been referred to
as the FIMP miracle~~\cite{McDonald:2001vt,Kang:2014cia,Kang:2015aqa}.

However, because of the relatively large DM self-interactions, the
situation is more complicated than originally studied in \cite{McDonald:2001vt,Kang:2014cia,Kang:2015aqa}.
Strong self-interactions may thermalize the hidden sector
within itself after the initial DM production from the SM sector has ended. By
thermalization we mean generation of not only kinetic but also chemical
equilibrium in the hidden DM sector, which tends to change the DM number
density even after the coupling between DM and the SM fields has been effectively
shut off. Then the final DM abundance
becomes determined by the dark freeze-out occurring in the hidden sector
instead of the usual freeze-in mechanism discussed above \cite{Carlson:1992fn,Chu:2011be,Bernal:2015ova,Bernal:2015xba,Heikinheimo:2016yds,Pappadopulo:2016pkp,Farina:2016llk,Bernal:2017mqb}.

Thus, the resulting abundance has to be calculated more carefully. 
In this work, we carry out a thorough analysis of DM
production in several benchmark scenarios where a scale invariant hidden sector contains
either a scalar, fermion (sterile neutrino), or vector DM candidate, and has sufficiently strong self-interactions. In the scenarios we study the observed DM abundance requires a small value for
the DM particle mass, compatible with large self-interaction
strength, which then leads to the final DM abundance
determined by the dark freeze-out after the non-thermal production from
the SM heat bath. As the model parameters are strongly constrained in order to
reproduce observations, and subtly related to the electroweak scale,
we call this the 'WIMP miracle of the second kind'.

The paper is organized as follows: in Section~\ref{miracle}, we
describe the models under investigation, review the original idea of the FIMP
miracle and discuss in detail the modifications needed in the
analysis in the presence of
sizeable self-interactions. Then, in sections~\ref{scalarDM}, \ref{fermionDM}, and \ref{vectorDM} we present our results for scalar, fermion, and vector dark
matter, respectively,
and finally conclude with a brief outlook in Section \ref{conclusions}.

\section{The FIMP miracle}
\label{miracle}

Consider the Lagrangian
\be
{\cal L} = {\cal L}_{\rm{SM}}+\frac12\partial_\mu s\partial^\mu s
-\frac{\lambda_{\rm s}}{4}s^4-V(\Phi,s),
\label{Eq:confLagrangian}
\ee
where ${\cal L}_{\rm{SM}}$ is the Lagrangian of the Standard Model,
$\Phi$ is the SM Higgs doublet, and $s$ is a real scalar field which is singlet under all SM interactions. The motivation for this form of the Lagrangian is to explain some beyond the SM phenomena, in our case the existence of
DM, without inputting a new ad hoc scale by hand and
subsequently generating a hierarchy problem for the electroweak scale
and the scale of new physics. There are no physical quadratic divergences in
the above Lagrangian and hence no hierarchy problem~\cite{Aoki:2012xs}.
Furthermore, the naturalness of the hidden sector is protected in the limit $\lambda_{\rm hs}\rightarrow 0$, where the hidden sector completely decouples from the SM~\cite{Foot:2013hna}. Thus the small portal coupling required for the freeze-in mechanism is technically natural in this setting. In this framework we have not provided a dynamical origin for the electroweak scale, and thus the naturalness problem of the SM persists. However, the scale invariant hidden sector is natural in isolation, and therefore, due to the smallness of the portal coupling, also protected from any naturalness problems originating from the SM sector.

The scalar potential is given by
\begin{equation}
\label{potential}
V(\Phi,s) = \frac{\lambda_{\rm hs}}{2}\Phi^\dagger\Phi s^2 ,
\end{equation}
where the SM Higgs doublet is assumed to obtain a vacuum expectation value
(vev), $\sqrt{2}\Phi = (0,v+h)$. Furthermore, we assume
$\lambda_{\rm s}>0$ for stability of the potential and
$\lambda_{\rm hs}>0$ in order not to induce a vev for the singlet
scalar $s$. Thus, after the SM
Higgs gains a vev, the singlet sector scale-invariance is spontaneously broken
and the singlet scalar gains a mass given by
\begin{equation}
m_{\rm s} = \sqrt{\frac{\lambda_{\rm hs}}{2}}v ,
\label{Eq:smass}
\end{equation}
where $v=246$ GeV is the Higgs vev. As we are considering a scenario where the
singlet $s$ is a FIMP, we note that the key element of the freeze-in
mechanism is the assumption that the portal coupling takes a very
small value, $\lambda_{\rm hs}\leq 10^{-7}$ \cite{Enqvist:2014zqa}, to avoid thermalization
of the hidden sector with the SM sector. Already this fixes $m_{\rm
s}\leq 0.1$ GeV in this scenario.

In the absence of large DM self-interactions, the yield is given by
~\cite{Hall:2009bx}
\begin{equation}
\label{DMyield}
\frac{\Omega_{\rm DM}h^2}{0.12} \simeq 8.3\times 10^{18}\lambda_{\rm
hs}^2\left(\frac{m_{\rm s}}{10\, {\rm MeV}}\right)
\simeq 1.4\times 10^{23}\lambda_{\rm hs}^{5/2} ,
\end{equation}
where, in the last step, we used Eq. \eqref{Eq:smass}. The yield arises from decays of the Higgs bosons during a short time interval between the moment when the Higgs field acquired a vacuum expectation value around $T\sim m_{\rm h}$, and the moment when their number density became Boltzmann-suppressed, $T\sim m_{\rm h}/3$. Requiring \eqref{DMyield} to match the observed relic density, we find
$\lambda_{\rm hs}\simeq 5\times 10^{-10}$, and thus also
$m_{\rm s}\simeq 4$ MeV.
In this way, the scale invariant FIMP model naturally leads to small
values for the dark matter mass, which in turn imply large DM
self-interactions for modest values of $\lambda_{\rm s}$.

Explicitly, because $m_{\rm s}\ll m_{\rm h}$, the singlet self-interaction strength is
\begin{eqnarray}
\frac{\sigma_{\rm s}}{m_{\rm s}} &=& \frac{9\lambda_{\rm s}^2}{32\pi m_{\rm s}^3} \\ \nonumber
&\simeq& \left(\frac{\lambda_{\rm hs}}{10^{-9}}\right)^{-3/2}\left(\frac{\lambda_{\rm s}}{0.1}\right)^2\frac{{\rm cm}^2}{{\rm g}} ,
\end{eqnarray}
which means that in the case of positive observation of $\sigma_{\rm s}/m_{\rm s}$, the result fixes also $\lambda_{\rm s}$. By taking $\sigma_{\rm s}/m_{\rm s}=1 {\rm cm}^2/{\rm g}$ as a representative example, we find that the scenario uniquely fixes the model parameters to\footnote{One may worry if such a large quartic scalar coupling will be driven to nonperturbatively large values at some scale $\mu\ll M_{\rm{P}}$. At one loop order, and for initial value $\lambda_{\rm{s}}< 0.1$
this will not happen.} $\lambda_{\rm s}\simeq 0.06$ and $\lambda_{\rm hs}\simeq 5\times 10^{-10}$, as above. This is the reason why the scenario has been referred to as the FIMP miracle.

However, it is not clear whether this large scalar self-interaction coupling is consistent with the assumptions made in deriving the DM abundance. Indeed, the result \eqref{DMyield} is applicable only in the limit of small self-interactions, where the comoving DM abundance does not change once produced from the SM sector. In the remaining of this paper, we discuss how the self-interactions are taken into account in different models in a consistent way, and what this means for the FIMP miracle.

\section{Production of dark matter}
\label{DMgenesis}

In this section we study consecutively three different scenarios, starting with the case where the hidden sector consists of a real singlet scalar $s$ only. Then we extend the model to include also a sterile neutrino and as a third case we study a scenario where the singlet scalar is promoted to be a complex doublet of a hidden $SU(2)_{\rm D}$ gauge symmetry.

\subsection{Scalar dark matter}
\label{scalarDM}

Let us start with the benchmark scenario where the hidden sector consists only of a real singlet scalar $s$, whose potential is given by \eqref{potential} and which acquires a mass given by \eqref{Eq:smass} once the electroweak symmetry breaks. In case of substantially large self-interactions, the final DM abundance is not given by the usual freeze-in mechanism but by a dark freeze-out,
operating in the hidden sector.

\begin{figure}
\begin{center}
\includegraphics[width=.5\textwidth]{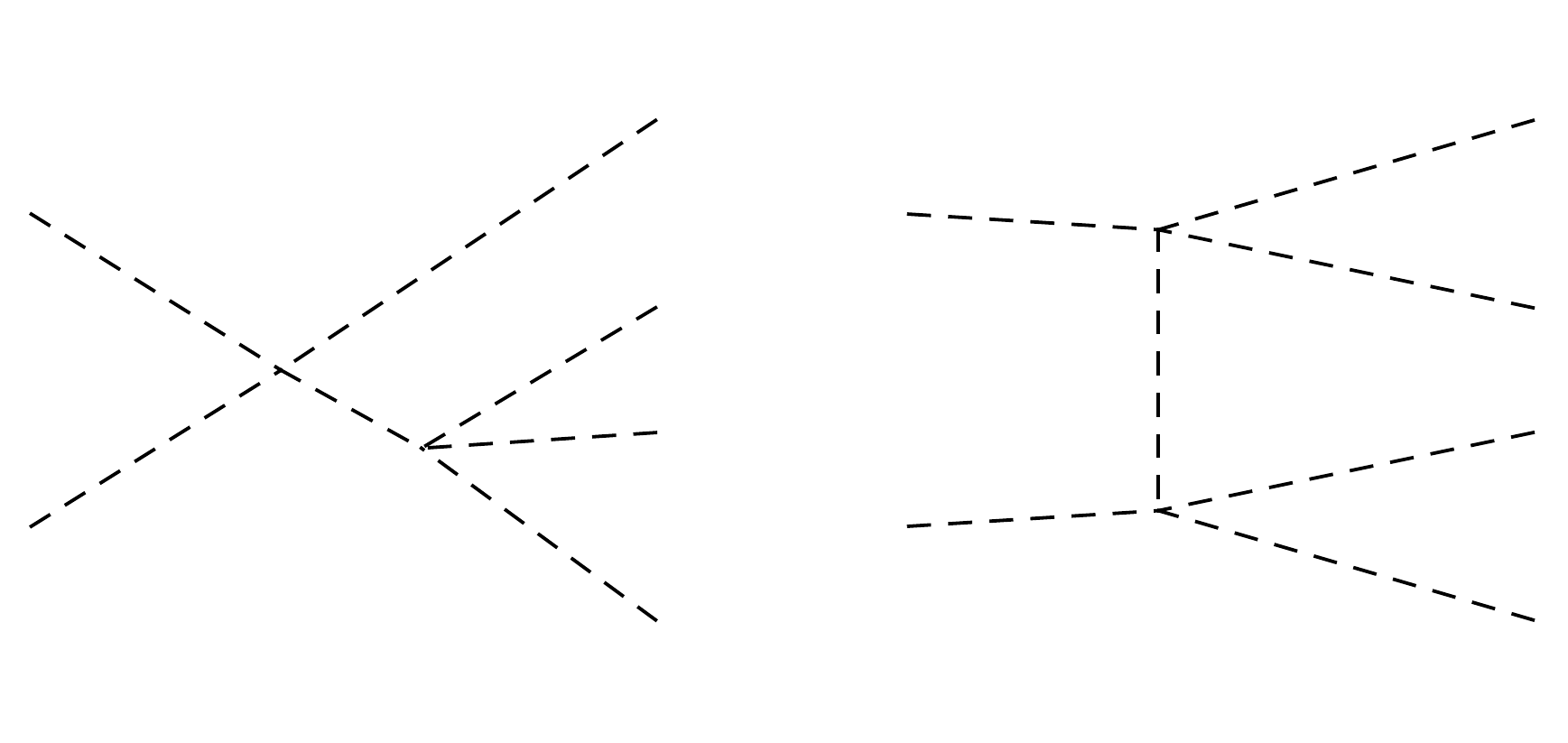}
\caption{Examples of Feynman diagrams for the $2\to 4$ scalar self scattering process.}
\label{diagram}
\end{center}
\end{figure}

The thermal history of the hidden sector proceeds in this case as follows: An initial abundance of $s$ particles is produced through Higgs decays
~\cite{Chu:2011be}
\be
n_{\rm D}^{\rm initial} \simeq \left. 3\frac{n_{\rm h}^{\rm eq}\Gamma_{h\rightarrow ss}}{H}\right|_{T=m_{\rm h}},
\label{eq:n init}
\ee
where $n_{\rm h}^{\rm eq}$ is the equilibrium number density of the Higgs boson, $H$ is the Hubble rate and the expression is evaluated when the temperature of the SM plasma is $T\approx m_{\rm h}$. The Higgs decay width into $s$ particles is
\be
\Gamma_{h\rightarrow ss} = \frac{\lambda_{\rm hs}^2 v_{\rm EW}^2}{32\pi m_{\rm h}}.
\label{eq:htoss}
\ee
If number changing interactions, i.e. the $2\rightarrow 4$ scattering processes in the hidden sector are fast, they will lead to a chemical equilibrium within the hidden sector, reducing the average momentum of $s$ particles and increasing their number density; see Fig. \ref{diagram} for clarity. This happens if the magnitude of $\lambda_{\rm s}$ exceeds the critical value ~\cite{Heikinheimo:2016yds}
\be
\lambda_{\rm s}^{\rm crit} \simeq \sqrt{\frac{52.7(g_*(m_{\rm h})g_*(m_{\rm s}/\xi))^\frac14 \sqrt{m_{\rm h} m_{\rm s}}}{\lambda_{\rm hs}M_{\rm P} }},
\label{lambda crit scalarDM}
\ee
where $g_*(T)$ is the effective number of relativistic degrees of freedom in the SM plasma at temperature $T$, $M_{\rm P}$ is the Planck mass and $\xi \equiv T_{\rm D}/T = (g_*\rho_{\rm D}/(\rho g_{*{\rm D}}))^{1/4}$ is the ratio of dark sector and SM temperatures. Here $g_{*D}=1$ is the number of relativistic degrees of freedom in the hidden sector, $\rho$ is the energy density of the SM plasma, and \mbox{$\rho_{\rm D} \simeq \frac12 n_{\rm D}^{\rm initial}m_{\rm h}$} is the energy density of the hidden sector, where $\frac12 m_{\rm h}$ is the average energy of the DM particles created via Higgs decays. For simplicity we approximate the number of relativistic degrees of freedom in the SM plasma as $g_*(T)=100$ for $T\sim m_{\rm h}$, and $g_*(T)=10$ for $T\sim m_{\rm s}\sim 10\ {\rm MeV}$.

If $\lambda_{\rm s} > \lambda_{\rm s}^{\rm crit}$, the $2\leftrightarrow 4$ interactions then maintain the equilibrium number density
at the hidden sector temperature $T_{\rm D}$, until the $4\rightarrow 2$ interaction rate drops below the Hubble rate and the $s$ number density finally freezes out. The corresponding dark freeze-out temperature $x_{\rm D}^{\rm FO}\equiv m_{\rm s}/T_{\rm D}^{\rm FO}$ can be solved from the implicit equation \cite{Heikinheimo:2016yds}
\be
x_{\rm D}^{\rm FO} = \frac13 \log\left(\left(\frac{1}{2\pi}\right)^\frac92 \frac{\lambda_{\rm s}^4\xi^2 M_{\rm P}}{1.66\sqrt{g_*}m_{\rm s} (x_{\rm D}^{\rm FO})^\frac52 }\right).
\label{eq:xD2}
\ee

Since the hidden sector and the SM are thermally disconnected after the energy transfer to the hidden sector via Higgs decays has stopped, the entropy of both sectors is conserved separately after the singlet scalars have attained thermal equilibrium within themselves. As was first noticed in \cite{Carlson:1992fn}, the entropy conservation of the hidden sector can be used to express the present day DM abundance as a function of the dark freeze-out temperature as
\be
\Omega_{\rm DM}h^2 = \frac{g_{*D}}{g_*}\frac{\xi^3 m_{\rm s}}{x_{\rm D}^{\rm FO}3.6\times 10^{-9}\ {\rm GeV}}.
\ee

\begin{figure}
\begin{center}
\includegraphics[width=.75\textwidth]{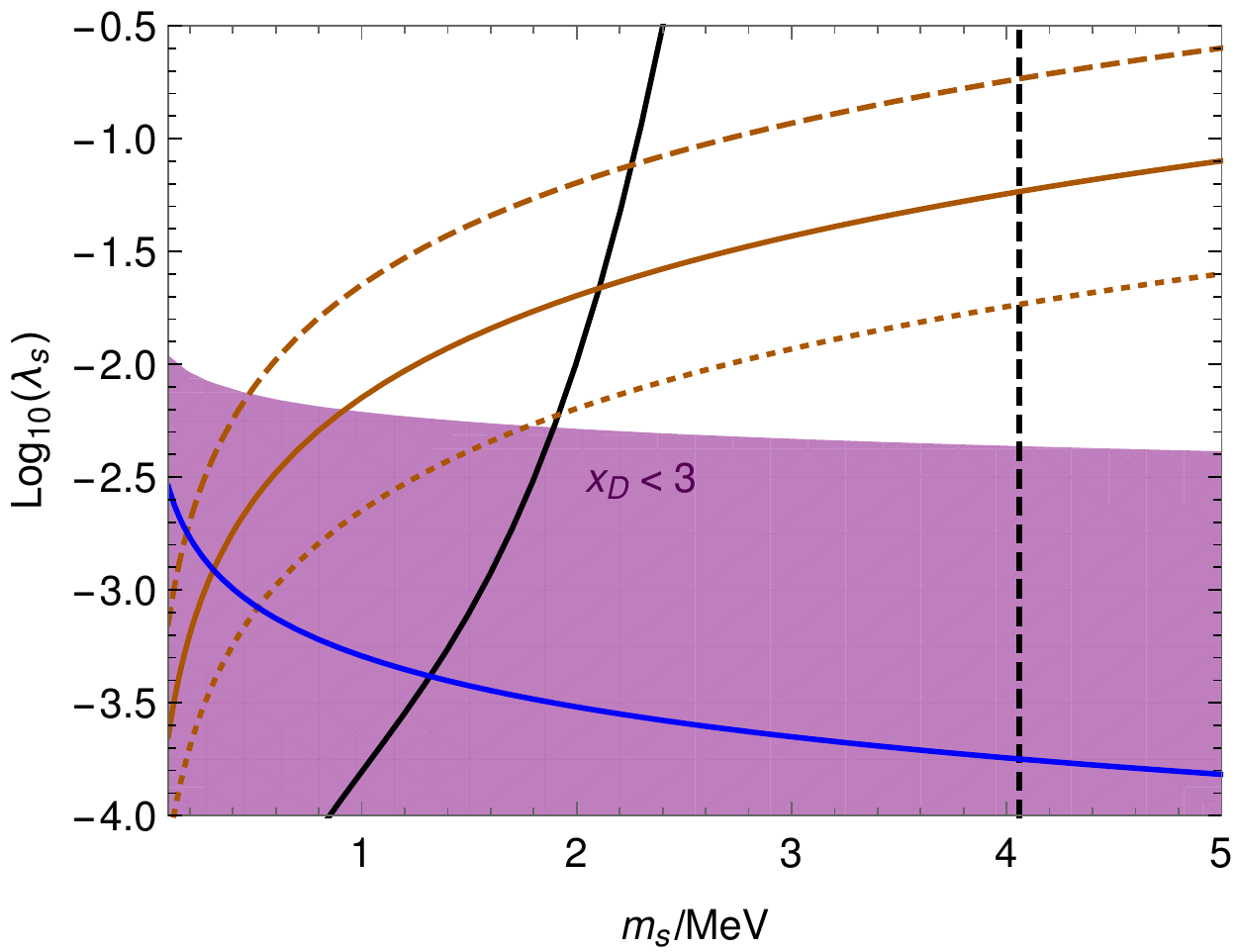}
\caption{Scalar dark matter abundance and self-interaction strengths in the $(m_{\rm s}, \lambda_{\rm s})$ space. Here the solid blue line is the limit for dark matter thermalization, so that above it the dark matter abundance (solid black line) is given by the dark freeze-out, whereas below the blue line the usual freeze-in mechanism is sufficient, and the observed abundance is produced for $m_{\rm s}\approx 4$ MeV, as depicted by the black dashed line. The dashed, solid, and dotted brown lines depict dark matter self-interactions $\sigma/m_{\rm s}=10,1,0.1\ {\rm cm}^2/{\rm g}$, respectively. In the purple shaded region the dark freeze-out happens while the DM is still relativistic ($x_{\rm D}^{\rm FO}<3$), so that the calculation of the abundance is subject to large relativistic corrections.}
\label{scalarDMplot}
\end{center}
\end{figure}

The result is depicted in Fig. \ref{scalarDMplot}. The figure shows how the 'WIMP miracle of the second kind' is possible for $m_{\rm s}\simeq 2$ MeV: the correct DM abundance, shown by the solid black line, is obtained simultaneously with $\sigma_{\rm s}/m_{\rm s}\simeq 1 {\rm cm}^2/{\rm g}$, shown by the brown contours. The DM yield from the usual freeze-in mechanism is shown by the vertical dashed black line. This mechanism is valid below the solid blue line, where the hidden sector does not thermalize. In case of a positive observation of a non-zero dark matter self-interaction strength of this order, the abundance and self-interaction strength uniquely fix the model parameters to $\lambda_{\rm hs}\simeq 1.5\times 10^{-10}$ and $\lambda_{\rm s}\simeq 0.02$. The original FIMP miracle calculation gives a result which is close to the one where the hidden sector thermalization is taken into account in a consistent way; the difference is less than an order of magnitude in the coupling values.

\subsection{Fermion dark matter}
\label{fermionDM}

We will next consider a slightly more complicated hidden sector, where a gauge-singlet fermion\footnote{The gauge-singlet fermion is essentially a sterile neutrino with a vanishing active--sterile neutrino mixing angle.} plays the role of dark matter and the scalar singlet discussed in the previous section takes the role of a messenger, mediating the interactions between the hidden and visible sectors via the Higgs portal coupling. We maintain the tree-level scale invariance in the hidden sector, so that the DM mass scale originates solely from the SM electroweak symmetry breaking scale.
With the notation introduced in the previous sections, the hidden sector Lagrangian is now given by

\begin{equation}
\mathcal{L}_{\rm hidden} = \frac12\partial_\mu s\partial^\mu s+i\bar{\psi}\slashed{\partial}\psi+\frac{\lambda_{\rm s}}{4}s^4 + ys\bar{\psi}\psi-V(\Phi,s) ,
\end{equation}
where we assume $\lambda_{\rm s}, y>0$, but take $\lambda_{\rm hs}<0$, so that the hidden sector scale invariance is spontaneously broken by the Higgs vacuum expectation value. Note that the value of the portal coupling is bounded from below by the scalar potential stability requirement $\lambda_{\rm hs} > -2\sqrt{\lambda_{\rm h}\lambda_{\rm s}}$. The singlet scalar $s$ then acquires a vev
\begin{equation}
\langle s\rangle = \sqrt{\frac{|\lambda_{\rm hs}|}{2\lambda_{\rm s}}}v ,
\end{equation}
and, as a result, $s$ and $\psi$ acquire masses
\begin{eqnarray}
m_{\rm s}^2 &=& 3\lambda_{\rm s}\langle s \rangle^2 - \frac{|\lambda_{\rm hs}|}{2}v^2 = |\lambda_{\rm hs}|v^2 \\ \nonumber
m_\psi &=& y\sqrt{\frac{|\lambda_{\rm hs}|}{2\lambda_{\rm s}}}v ,
\end{eqnarray}
where we have ignored the mixing between the singlet scalar and the Higgs. The corrections arising from the mixing are of the order $\mathcal{O}(\lambda_{\rm hs}^{3/2})$, and are therefore negligible in the feebly coupled limit $\lambda_{\rm hs}\ll 1$, which is our interest here.
If $m_{\rm s}>2m_\psi$, i.e. $\lambda_{\rm s}>2y^2$, the singlet scalar can decay into fermions. In any case, $m_{\rm s}, m_\psi \ll m_{\rm h}$.

The $2\rightarrow 2$ self-interactions of the fermions are mediated by the scalar. The relevant quantity for observations regarding structure formation and mergers of galaxies or clusters is the so-called viscosity cross section\footnote{This is technically true also for the scalar DM scenario discussed above, but since the scalar self-scattering is a contact interaction and the resulting differential cross section does not depend on the scattering angle, the viscosity cross section is in that case simply related to the standard cross section by $\sigma_V=\frac34 \sigma$.}~\cite{Tulin:2013teo}
\begin{equation}
\sigma_V = \int d\Omega \sin^2\theta \frac{d\sigma_{\rm SI}}{d\Omega},
\end{equation}
where $\sigma_{\rm SI}$ is the $2\rightarrow 2$ elastic scattering cross section. In the non-relativistic limit the viscosity cross section is given by
\begin{equation}
\sigma_{V0} = \frac{4 y^4 m_\psi^2}{3\pi m_{\rm s}^4} .
\end{equation}
However, if the scalar is light, the elastic scattering cross section is enhanced by the Sommerfeld effect at low velocities. In this case the cross section can be approximated as $\sigma_V = \sigma_{V0}S(v)$, where $S(v)$ is the Sommerfeld enhancement factor estimated from the Hulthen potential~\cite{Blum:2016nrz}
\begin{equation}
S(v) = \frac{2\pi\alpha}{v}\frac{\sinh\left(\frac{2\pi\hat{p}}{\delta}\right)}{\cosh\left(\frac{2\pi\hat{p}}{\delta}\right)-\cos\left(2\pi\frac{\sqrt{\delta-\hat{p}^2}}{\delta}\right)},
\label{Sommerfeld}
\end{equation}
where $\alpha=y^2/4\pi$, $\hat{p} = p/(\alpha m_\psi)$, $\delta = m_*/(\alpha m_\psi)$ and $m_* = m_{\rm s}\pi^2/6$.

Generally, there are various possibilities for producing the abundance of the
fermion dark matter. For example, if $m_{\rm s} > 2 m_\psi$, the singlet
scalars could be produced via freeze-in, and they would eventually decay into
the fermions, yielding $n_\psi = 2n_{\rm s}$, where $n_{\rm s}$ is the initial
abundance of scalars given in \eqref{eq:n init}. However, this scenario is
inconsistent, since we require large
fermion self-interactions which will thermalize the hidden sector, and the final fermion abundance will be
determined via a freeze-out occurring in the dark sector.

In~\cite{Heikinheimo:2016yds} we studied the scenario where the scalars quickly
decay into fermions, and the fermion abundance is determined via the freeze-out
of the $4\rightarrow2$ fermion self-scattering. Here we find that due to the
assumed scale invariance of the hidden sector, the picture gets modified. If
$m_{\rm s} > 2 m_\psi$, i.e. $\lambda_{\rm s}>2y^2$, then the decay of the
scalars and $2\rightarrow 3$ scalar scattering occur over similar timescale and
the hidden sector becomes populated by both the scalars and fermions. The final DM abundance is determined by the freeze-out of the fermion
$4\rightarrow2$ scattering process, which takes place after the heavier scalars
have decoupled from equilibrium and decayed into fermions.

On the other hand, if we assume the opposite ordering of masses,
$2m_\psi > m_{\rm s}$, and that the hidden sector thermalizes so that both the fermions and scalars are in chemical equilibrium, the final abundance of fermions will be set by the freeze-out of the $\bar{\psi}\psi\rightarrow ss$ annihilations. The remaining scalars will then decay into SM particles, since for this mass hierarchy there are no kinematically allowed decay modes of the scalars inside the hidden sector.

We will consider in more detail the two possibilities outlined above. We start with the ordering where the scalar is the lightest state in the spectrum.

\subsubsection{Mass hierarchy $m_\psi > m_{\rm s}$}
\label{fermionsub2}

In this case, we must first require that the hidden sector reaches chemical equilibrium, while the equilibrium temperature $T_{\rm D}$ is still above the fermion mass. Otherwise the fermions will not necessarily reach their equilibrium number density, and the computation of their abundance via freeze-out cannot be justified. The thermalization may happen via any of the possible scattering processes that change the total particle number $n_{\rm s}+n_\psi$ within the hidden sector. For simplicity, here we consider only the process $ss\rightarrow \bar{\psi}\psi s$, which is dominant in the limit $y^2\gg \lambda_{\rm s}$. We approximate the velocity averaged scattering cross section of this process by
\be
\langle \sigma_{ss\rightarrow \bar{\psi}\psi s} v\rangle \approx \frac{y^6}{p^2},
\ee
where $p$ is the four-momentum of the incoming scalars in the center of mass frame. The scalars are initially produced from Higgs decays with $p_0^2= m_{\rm h}^2/4$, and the momentum is then redshifted as $p(T)=p_0 T/T_0$, where $T_0\approx m_{\rm h}$ is the temperature of the SM photon bath when the energy transfer from the SM to the hidden sector stops. We require that the hidden sector reaches chemical equilibrium before the equilibrium temperature $T_{\rm D}$ drops below the fermion mass, {\it i.e.} we require \mbox{$n_{\rm s} \langle \sigma_{2\rightarrow 3} v\rangle > H$} at $T_{\rm D} = m_\psi$, where $n_{\rm s}$ is the initial number density of the scalars \eqref{eq:n init} diluted with the scale factor as $(a_0/a)^3$ down from $T_0$. This can be cast as a lower limit for the Yukawa coupling as
\be
y^{\rm crit} \approx 3\left(\frac{\sqrt{g_*(m_{\rm h})g_*(m_\psi/\xi)}m_{\rm h}^3m_\psi}{\lambda_{\rm hs}^2\xi v^2 M_{\rm P}^2}\right)^\frac16 ,
\label{gcrit}
\ee
where $\xi=T_{\rm D}/T$. For $y > y^{\rm crit}$ the hidden sector reaches a chemical equilibrium and the final abundance of fermions is determined via freeze-out of the $\bar{\psi}\psi\rightarrow ss$ annihilation process as~\cite{Chu:2011be}
\begin{equation}
\Omega_{\rm DM} h^2 \approx 2\frac{1.07\times 10^9\ {\rm GeV}^{-1}\xi^2 x_f^2}{\sqrt{g_*}M_{\rm P}\sigma_0^{\rm ann}},
\end{equation}
where the freeze-out temperature $x_{\rm D}^{\rm FO} = m_\psi/T_D^{\rm FO}$ is given by
\begin{equation}
x_{\rm D}^{\rm FO} = \log\left(\xi^2\frac{M_{\rm P}m_\psi\sigma_0^{\rm ann}}{1.66\sqrt{g_*}(2\pi)^\frac32 \sqrt{x_{\rm D}^{\rm FO}}}\right),
\end{equation}
where the thermally averaged cross section of the $p$-wave annihilation process is \mbox{$\langle \sigma v\rangle \approx\sigma_0^{\rm ann}(x_{\rm D}^{\rm FO})^{-1}$}, and in the nonrelativistic limit for $m_{\rm s} \ll m_\psi$ is given by
\begin{equation}
\sigma_0^{\rm ann} \approx \frac{3y^4}{16\pi m_\psi^2}.
\end{equation}
After the freeze-out of the fermion DM abundance, the hidden sector still contains a population of the scalars $s$, which will eventually decay into the SM. Due to the smallness of the portal coupling, their lifetime is potentially long, so that they may deposit energy into the SM photon bath during of after Big Bang Nucleosynthesis (BBN), potentially destroying the success of the SM prediction for abundances of light elements produced during BBN. The scalar couples to the SM primarily via the mixing with the Higgs, where the mixing angle is given by
\begin{equation}
\sin\theta_{\rm mix} \approx \frac{\lambda_{\rm hs}^{3/2}}{\lambda_{\rm h}\sqrt{\lambda_{\rm s}}},
\end{equation}
where $\lambda_{\rm h}\approx 0.13$ is the SM Higgs quartic coupling. The width of the dominant diphoton decay channel for a light $s$ is then approximated as
\begin{equation}
\Gamma_{s\rightarrow \gamma\gamma} \approx \frac{m_{\rm s}^3}{m_{\rm h}^3}\sin^2\theta_{\rm mix}\Gamma_{h\rightarrow \gamma\gamma},
\end{equation}
where $\Gamma_{h\rightarrow \gamma\gamma} \approx 9.28\times 10^{-6}\ {\rm GeV}$ is the Higgs diphoton decay width. As a rough estimate, in order to evade the constraints from light element abundancies, we require that the scalar lifetime is below $\tau_{\rm s} < 10^5\ {\rm s}$~\cite{Pospelov:2010hj,Poulin:2016anj}.

\begin{figure}
\begin{center}
\includegraphics[width = 0.6\textwidth]{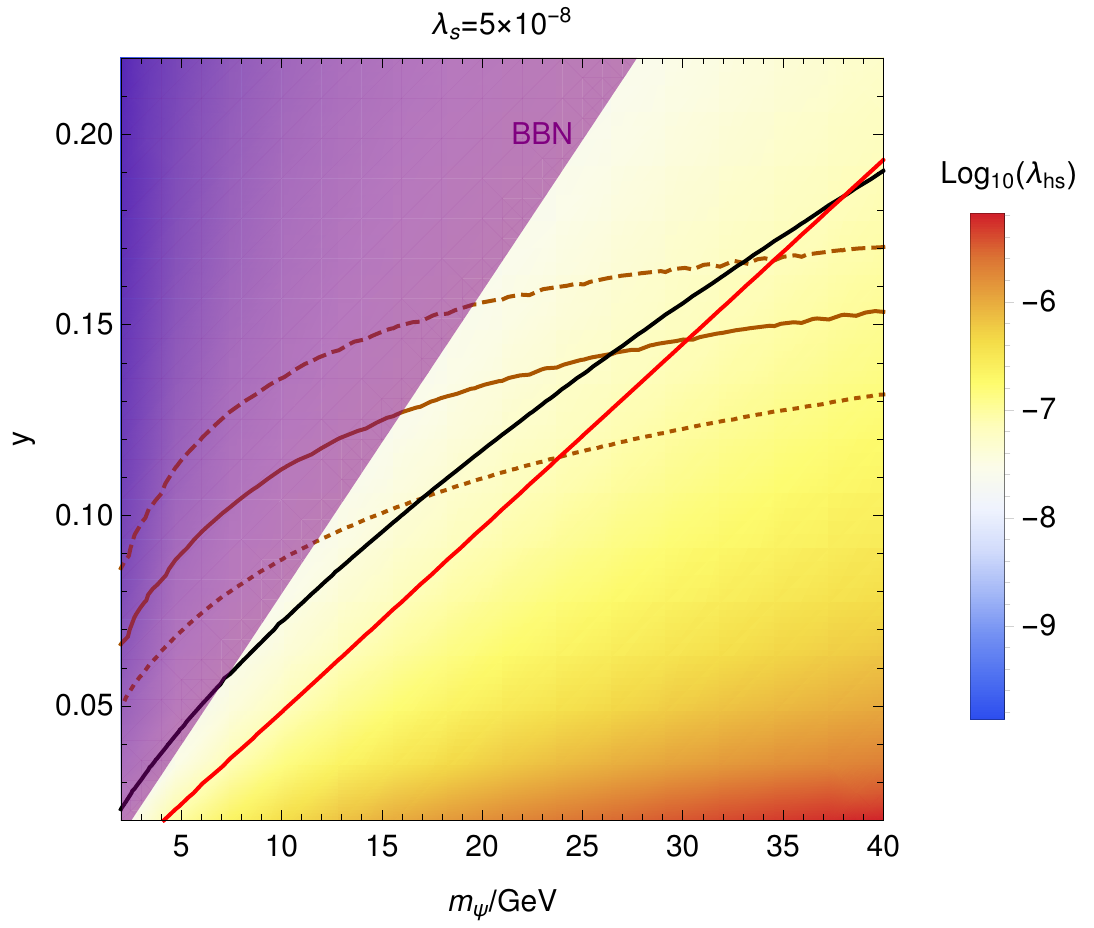}
\caption{The correct fermion DM abundance is produced along the black solid line via the freeze-out of the $\bar{\psi}\psi\rightarrow ss$ process, following the thermalization of the hidden sector via $ss\rightarrow \bar{\psi}\psi s$. The scalar quartic coupling has been fixed as $\lambda_{\rm s} = 5\times 10^{-8}$. The dotted, solid and dashed brown lines show the DM self-interaction cross section \mbox{$\sigma_V/m_\psi = 0.1,1,10\ {\rm cm}^2/{\rm g}$}, respectively, evaluated at the typical velocity scale for dwarf halos, $v\sim 10$ km/s~\cite{Tulin:2013teo}. The color code shows the value of the portal coupling $\lambda_{\rm hs}$ on a logarithmic scale. Below the red contour the hidden sector reaches thermal equilibrium with the SM, and the freeze-in picture is not consistent. The purple shaded region shows the BBN constraint for the scalar lifetime.}
\label{fermionDMplot2}
\end{center}
\end{figure}

The results of this analysis are shown in Fig. \ref{fermionDMplot2}. We find that the correct DM abundance can be produced via the dark freeze-out mechanism while simultaneously obtaining DM self-interaction cross section over mass in the region of interest between 0.1 and 10 ${\rm cm}^2/{\rm g}$ for the scalar self coupling $\lambda_{\rm s}\lesssim 10^{-7}$, with $y\sim 0.1$ and $m_\psi\sim 20\ {\rm GeV}$. For larger values of the scalar quartic coupling the self-interacting region moves below the red contour in the plot, corresponding to the thermalization between the hidden sector and the SM, and loss of the initial freeze-in mechanism. For smaller values of the quartic coupling the lifetime of the scalar exceeds the BBN bound $\tau_{\rm s} < 10^5\ {\rm s}$.

However, there is a caveat to this scenario that presumably renders it phenomenologically unacceptable. In the region of interest that would result in the observed DM abundance, the mass of the scalar is of the order of $m_{\rm s}\sim 10\ {\rm MeV}$, and the lifetime is close to the BBN bound, $\tau_{\rm s}\lesssim 10^{5}\ {\rm s}$. Thus, before decaying into the SM, the scalars will have cooled down enough to become nonrelativistic. At this point, the energy density stored in the scalars begins to scale as $\sim a^{-3}$ as a function of the scale factor, whereas the radiation dominated SM bath scales as $\sim a^{-4}$. Thus the nonrelativistic scalars will soon begin to dominate the energy density of the universe, resulting in a non-standard epoch of matter domination between BBN and recombination. We have not explored in detail the consequences of this scenario, but it is likely to destroy the phenomenological success of the standard $\Lambda$CDM model in predicting the light element abundances and the CMB power spectrum.

\subsubsection{Mass hierarchy $m_{\rm s} > 2m_\psi$}
\label{fermionDMsub1}

Similarly to the case for the scalar dark matter discussed in Section \ref{scalarDM}, the hidden sector will reach chemical equilibrium if the number changing scalar self scattering rate exceeds the Hubble rate. The difference to the scalar DM scenario is that now the scalar $\mathbb{Z}_2$ symmetry is broken by the scalar vacuum expectation value, and thus the leading number changing interaction is the $2\rightarrow 3$ process. We approximate the velocity averaged cross section of this process as
\be
\langle \sigma_{ss\rightarrow sss}v\rangle \approx \frac{\lambda_{\rm s}^4 v_{\rm s}^2}{p^2 m_{\rm s}^2}\approx\frac{\lambda_{\rm s}^3}{p^2}.
\ee
This yields the critical coupling for thermalization of the Hidden sector as
\be
\lambda_{\rm s}^{\rm crit} \approx 16\left(\frac{\sqrt{g_*(m_{\rm h})g_*(m_{\rm s}/\xi)}m_{\rm h}^3m_{\rm s}}{\lambda_{\rm hs}^2 v^2 M_{\rm P}^2}\right)^\frac13\approx10^{-11}\lambda_{\rm hs}^{-\frac23}\left(\frac{m_{\rm s}}{{\rm GeV}}\right)^\frac13,
\label{lambda crit fermionDM}
\ee
in analogy to equation (\ref{lambda crit scalarDM}). 

The final DM abundance is then determined by the freeze-out of the fermion
$4\rightarrow2$ scattering process, taking place after the heavier scalars
have decoupled from equilibrium and decayed into fermions. The freeze-out temperature of the $4\rightarrow 2$ fermion self scattering process is given as
\be
x_{\rm D}^{\rm FO} = \frac13\log\left(\left(\frac{1}{2\pi}\right)^{\frac92}\frac{y^8m_\psi^9 M_{\rm P}}{1.66\sqrt{g_*}m_{\rm s}^{10}(x_{\rm D}^{\rm FO})^\frac52}\right),
\ee
where we have approximated the scalar-mediated fermion $4\rightarrow2$ self scattering cross section in the non-relativistic limit as
\be
\langle \sigma_{4\rightarrow2}v^3\rangle \approx \frac{y^8m_\psi^2}{m_{\rm s}^{10}}.
\label{fermion4to2}
\ee

\begin{figure}
\begin{center}
\includegraphics[width = 0.49\textwidth]{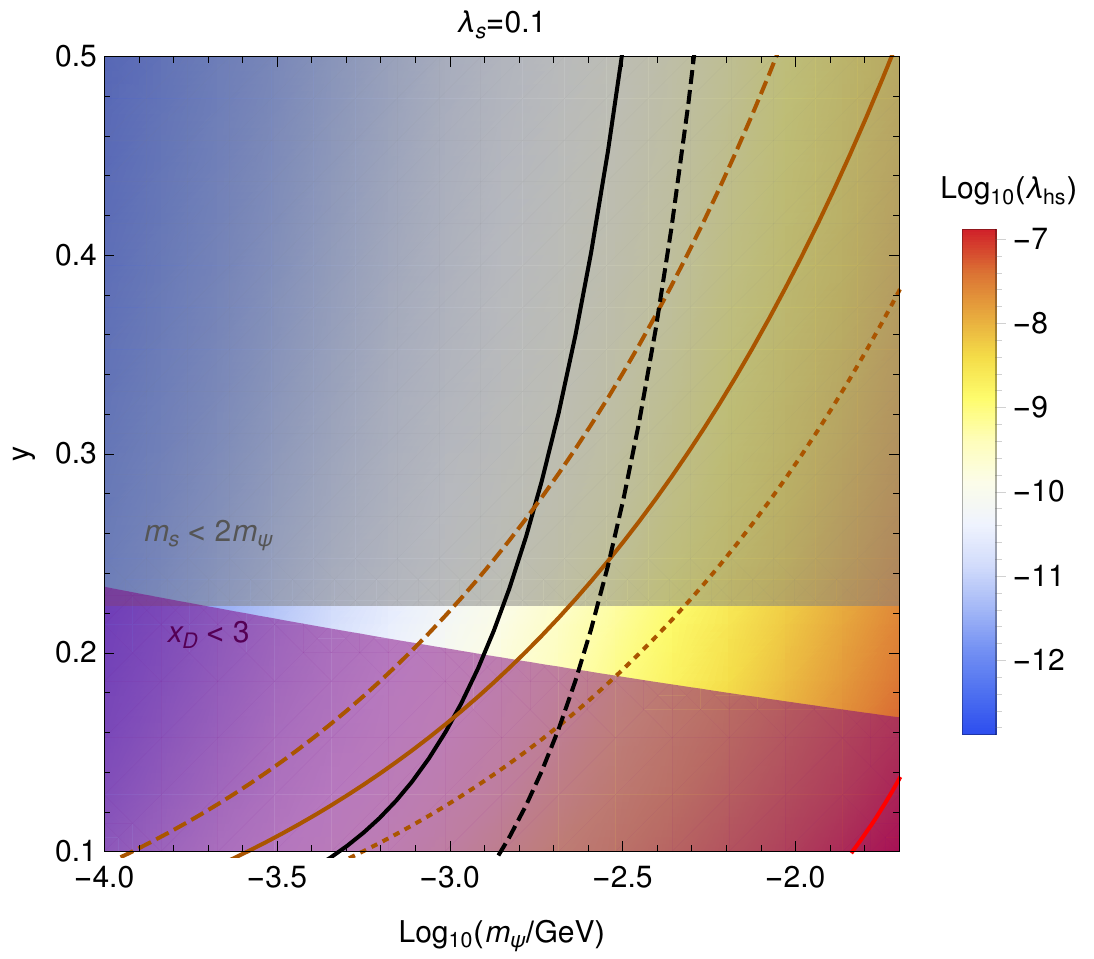}
\includegraphics[width = 0.49\textwidth]{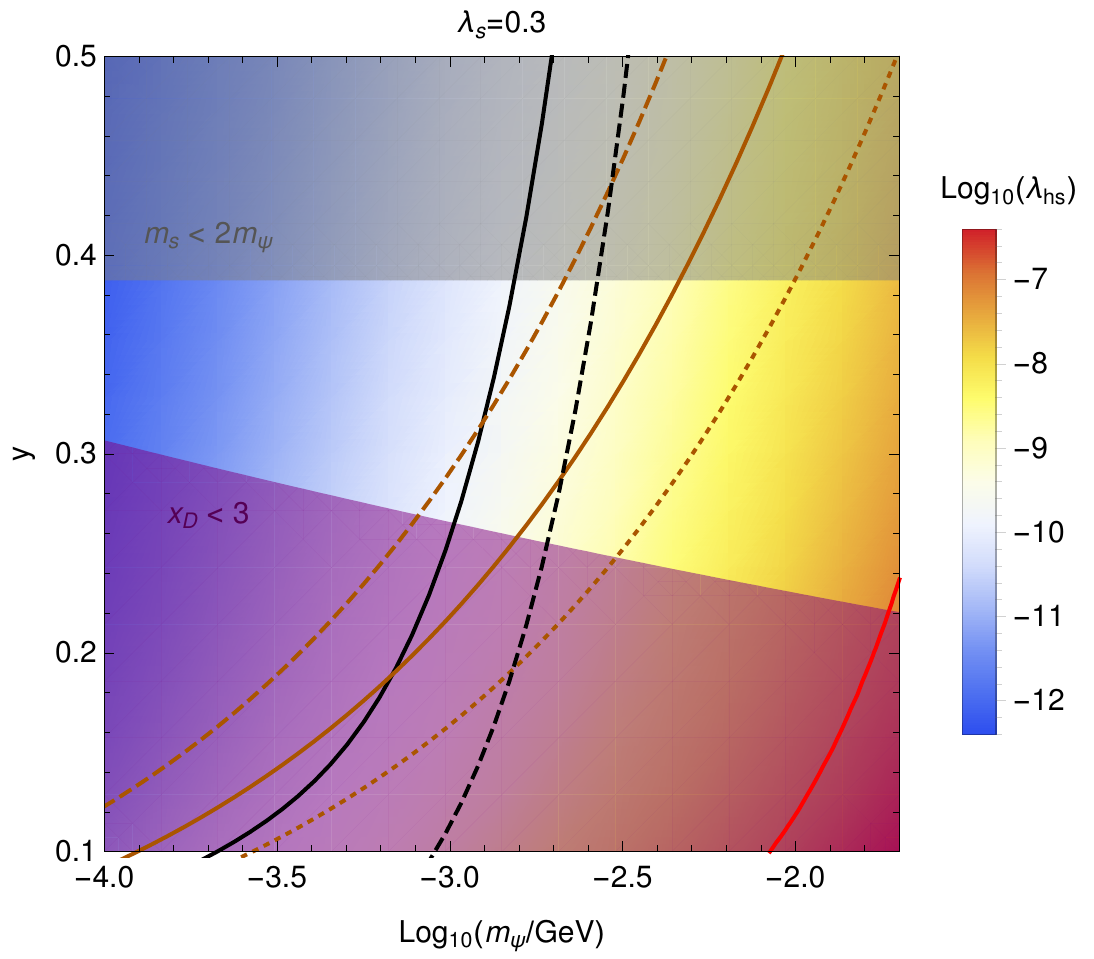}
\caption{The correct fermion DM abundance is produced along the black solid line via the freeze-out of the $\bar{\psi}\psi\bar{\psi}\psi\rightarrow \bar{\psi}\psi$ process, following the thermalization of the hidden sector via $ss\rightarrow sss$. The scalar quartic coupling has been fixed as $\lambda_{\rm s} = 0.1,\ (0.3)$ in the left (right) panel. The dotted, solid and dashed brown lines show the DM self-interaction cross section \mbox{$\sigma_V/m_\psi = 0.1,1,10\ {\rm cm}^2/{\rm g}$}, respectively. The color code shows the value of the portal coupling $\lambda_{\rm hs}$ on a logarithmic scale. Below the red contour the hidden sector reaches thermal equilibrium with the SM, and the freeze-in picture is not consistent. In the purple shaded region the freeze-out happens at a relativistic temperature $x_D < 3$, so that the calculation for the abundance is subject to large relativistic corrections. In the gray shaded region the assumption of the mass hierarchy $m_{\rm s} > 2m_\psi$ is not valid. The black dashed line shows for reference where the correct abundance would be obtained via the usual FIMP miracle calculation; see text for further discussion.}
\label{fermionDMplot}
\end{center}
\end{figure}

We find that the correct DM abundance can be produced via this mechanism in a small window of parameter space with the scalar self coupling obtaining values in the range $0.1\lesssim \lambda_{\rm s}\lesssim 1$. The correct abundance is then produced for $m_\psi\sim {\rm MeV}$ and $y\sim 0.3$. This situation is depicted in Fig. \ref{fermionDMplot}, where the observed DM abundance is produced via the freeze-out of the fermion $4\rightarrow 2$ self scattering as shown the black solid line. The black dashed line shows for reference where the correct abundance would be obtained via freeze-in of the scalars followed by the decay $s\rightarrow \bar{\psi}\psi$, ignoring the number changing self-interactions in the hidden sector. The dotted, solid and dashed brown lines show the DM self-interaction cross section \mbox{$\sigma_V/m_\psi = 0.1,1,10\ {\rm cm}^2/{\rm g}$}, respectively. 

We observe that in order to produce the correct DM abundance, the DM is necessarily self-interacting in this scenario. This is because the $y^8$ behaviour of the cross section (\ref{fermion4to2}) which, along with the large suppression from the number density for $4\rightarrow 2$ processes in general in the non-relativistic limit, tends to drive the freeze-out of the number changing processes to happen very early unless $y$ is large, thus resulting in overabundance of DM for small $y$. Furthermore, we observe that for $y\lesssim 0.2$ (slightly depending on $\lambda_{\rm s}$ and $m_\psi$) the number changing processes will freeze out while the fermions are still relativistic, where we have used the criterion $x_{\rm D}^{\rm FO}<3$, shown by the purple shaded region in the figure. In this region the result shown by the black solid line is subject to large relativistic corrections.

Finally, we need to check that the values of the couplings quoted above remain perturbative up to large energies. Since the portal coupling
$\lambda_{\rm hs}$ is negligible, the running of $\lambda_{\rm s}$ and $y$ is
governed at one loop order by
\bea
16\pi^2\beta_{\lambda_{\rm s}} &=& 18\lambda_{\rm s}^2+8\lambda_{\rm s} y^2-8y^4,\nonumber \\
16\pi^2\beta_y &=& 5y^3.
\eea
Given the initial values $y\simeq 0.2$ and $\lambda_{\rm s}\simeq 0.1$ corresponding
to the left panel of Fig.~\ref{fermionDMplot},
we find that the couplings remain perturbative when evolving up to scales
20 orders of magnitude larger than the initial one. As the value of the scalar self interaction $\lambda_{\rm s}$ at scale $\mu_0$
is increased to 0.3, corresponding to the right panel of
Fig.~\ref{fermionDMplot}, and taking $y\simeq 0.2$ we find that $\lambda_{\rm s}$
hits a Landau pole at $\mu\simeq 10^{13}\mu_0$ in one loop analysis.

\subsection{Vector dark matter}
\label{vectorDM}

Finally, we will consider a scenario where the DM particle is a spin-1 vector boson. As a representative example we study the hidden vector dark matter model of \cite{Hambye:2008bq}, by promoting the scalar $s$ to be a complex doublet of a hidden $SU(2)_{\rm D}$ gauge symmetry, while still keeping it as a singlet under the SM gauge groups. For simplicity, we assume that the hidden sector fermion fields are absent in this case. The hidden sector Lagrangian is then

\begin{equation}
\mathcal{L}_{\rm hidden} =  \frac{1}{4}F^{\mu\nu}F_{\mu\nu} + (D^{\mu}s)^{\dagger}(D_{\mu}s) - \lambda_{\rm s}(s^{\dagger}s)^2 ,
\end{equation}
where $D_{\mu}s = \partial_{\mu}s -i\frac{g}{2}\tau^a A^a_\mu s$ with $A^a_\mu$ the $SU(2)_{\rm D}$ gauge fields, $\tau^a$ the Pauli matrices,  and we have again imposed classical scale invariance on the hidden sector, discarding any operators with dimensionful coefficients in the scalar potential present in the original model of \cite{Hambye:2008bq}.

The sign of the portal coupling $\lambda_{\rm hs}$ is again taken negative, so that $s$ acquires a vacuum expectation value from the electroweak symmetry breaking. As a result, the hidden sector scale invariance and the $SU(2)_{\rm D}$ gauge symmetry are broken and the gauge boson $A$ acquires a mass

\be
m_{\rm A} = \frac{g}{2}\sqrt{\frac{|\lambda_{\rm hs}|}{2\lambda_{\rm s}}}v ,
\ee
similarly to the fermion mass in the scenario discussed in Section \ref{fermionDM}. As discussed in \cite{Hambye:2008bq}, all three massive vector bosons are degenerate in mass and stable due to a custodial global $SO(3)$ symmetry of the hidden sector, which is not broken by the portal coupling.

The main difference to the fermion scenario is that due to the non-Abelian gauge structure of the hidden sector, the DM fields $A$ have self-interaction terms of their own, besides the one mediated by the scalar $s$. We approximate the non-relativistic viscosity cross section by
\be
\sigma_{V0} = \frac{g^4 m_{\rm A}^2}{4\pi m_{\rm s}^4} + \frac{g^4}{4\pi m_{\rm A}^2},
\ee
where the first term, dominant in the limit $m_{\rm s} \ll m_{\rm A}$, arises from scalar exchange, and the second term, which is dominant in the opposite limit, from the three- and four-point interaction terms of the non-Abelian gauge bosons. We have neglected the interference term, which is only relevant in the small region of parameter space where the two contributions are of similar magnitude. In the low velocity limit for $m_{\rm s} \ll m_{\rm A}$, the total viscosity cross section is given by $\sigma_V = \sigma_{V0}S(v)$, where $S(v)$ is the Sommerfeld factor given by (\ref{Sommerfeld}).

The analysis parallels that of the fermion treated in detail in the previous section, and we will be brief on the details here. We will again study the two scenarios for producing the dark matter abundance, corresponding to the mass hierarchies $m_{\rm s} > 2m_{\rm A}$, where the DM abundance is determined by the freeze-out of the $3\rightarrow2$ self scattering process of the vector DM particle, or $m_{\rm A}>m_{\rm s}$, where the abundance is produced via freeze-out of the $AA\rightarrow ss$ annihilation process.

With the mass hierarchy $m_{\rm A}>m_{\rm s}$ 
there is a critical value $g^{\rm{crit}}$ above which the vectors will reach chemical equilibrium in the hidden sector, in analogy with Eq. (\ref{gcrit}).
After the vectors have reached chemical equilibrium, their final abundance is
determined by the freeze-out of the annihilation process $AA\rightarrow ss$.
However, as discussed in Section \ref{fermionsub2}, this scenario is plagued by
the epoch of early matter domination, when the long lived scalars become
nonrelativistic and begin to dominate the energy density of the universe
before decaying into relativistic SM species.

\begin{figure}
\begin{center}
\includegraphics[width = 0.49\textwidth]{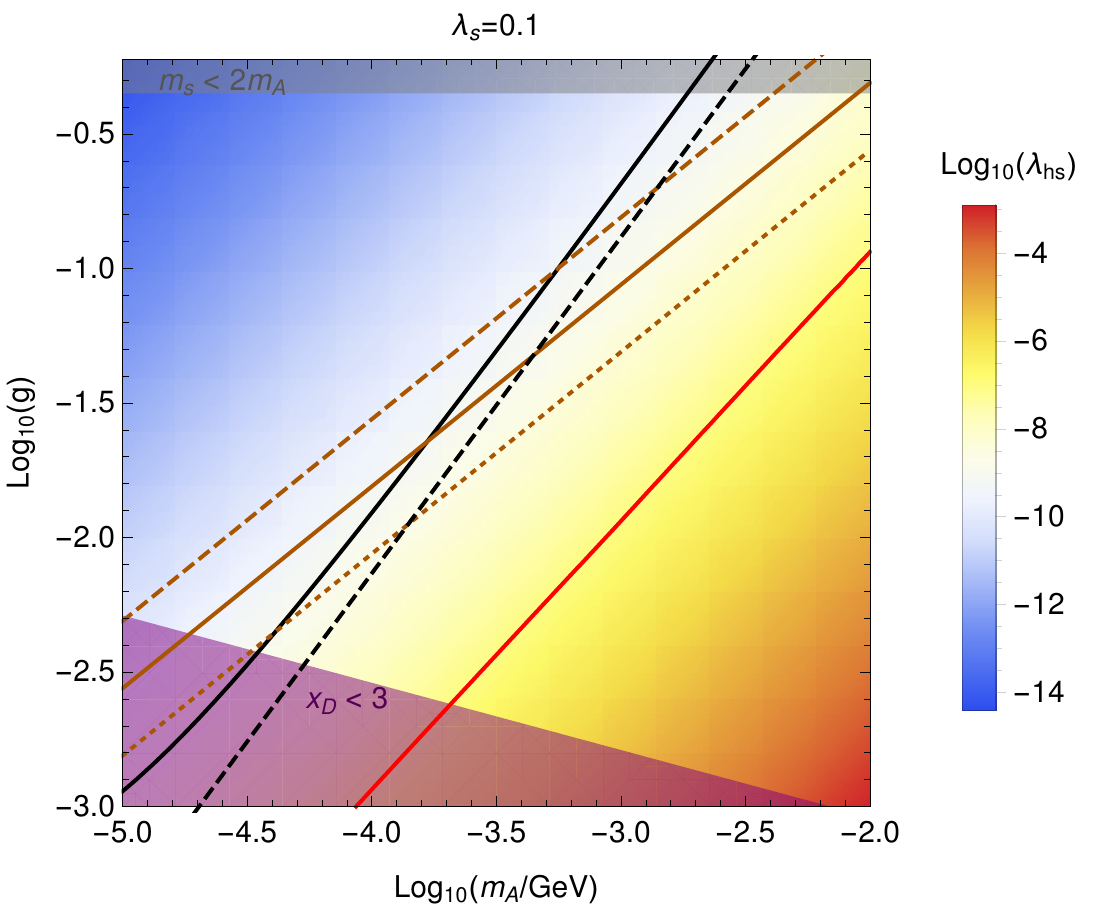}
\includegraphics[width = 0.49\textwidth]{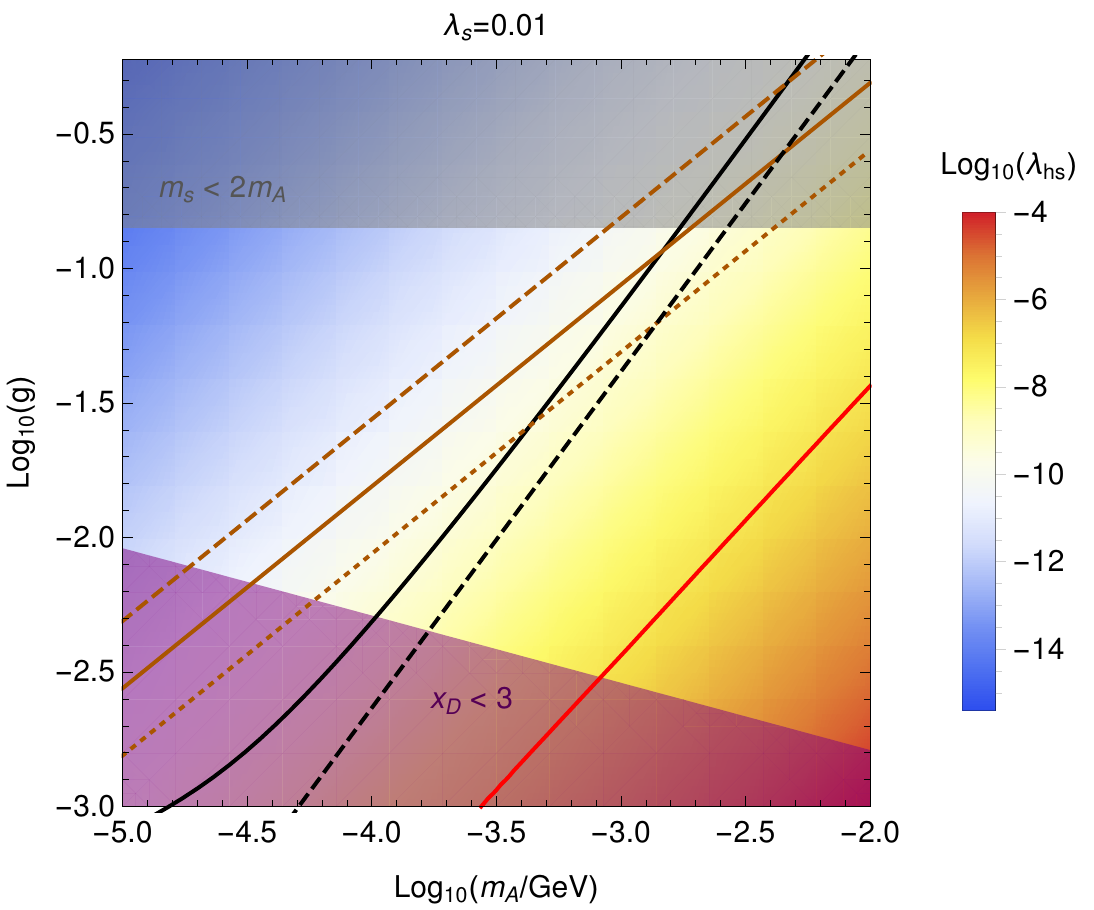}
\caption{The correct vector DM abundance is produced along the black solid line via the freeze-out of the $AAA\rightarrow AA$ process. The scalar quartic coupling has been fixed as $\lambda_{\rm s} = 0.1\ (0.01)$ in the left (right) panel. The dotted, solid and dashed brown lines show the DM self-interaction cross section \mbox{$\sigma_V/m_{\rm A} = 0.1,1,10\ {\rm cm}^2/{\rm g}$}, respectively. The color code shows the value of the portal coupling $\lambda_{\rm hs}$ on a logarithmic scale. Below the red contour the hidden sector reaches thermal equilibrium with the SM, and the freeze-in picture is not consistent. In the purple shaded region the freeze-out happens at a relativistic temperature $x_D < 3$, so that the calculation for the abundance is subject to large relativistic corrections, and in the gray shaded region the assumption of the mass hierarchy $m_{\rm s} > 2m_{\rm A}$ is not valid. The black dashed line shows for reference where the correct abundance would be obtained via the usual FIMP miracle calculation; see text for further discussion.}
\label{vectorDMplot}
\end{center}
\end{figure}

On the other hand, if $m_{\rm s} > 2m_{\rm A}$, thermalization of the hidden sector happens as in section \ref{fermionDMsub1}, via the scalar $2\rightarrow 3$ scattering, as described by Eq. (\ref{lambda crit fermionDM}). The hidden sector will then be populated by a thermal bath of scalars and vectors, until the heavier scalars freeze out and decay into vectors. The resulting bath of vector DM particles then remains in chemical equilibrium via the number changing self scattering processes, until the freeze-out of the $3\rightarrow2$ process at temperature $x_{\rm D}^{\rm FO} = m_{\rm A}/T_{\rm D}$ given by
\be
x_{\rm D}^{\rm FO} = \log\left(\xi^2\frac{M_{\rm P}m_{\rm A}^4\langle\sigma_{3\rightarrow 2}v^2\rangle}{(2\pi)^3 1.66\sqrt{g_*}x_{\rm D}^{\rm FO}}\right),
\ee
where we approximate the number changing self scattering cross section as
\be
\langle\sigma_{3\rightarrow 2}v^2\rangle\approx \frac{g^6}{m_{\rm A}^5}.
\ee
We find that the vector scenario, thanks to the less suppressed $3\rightarrow2$ number changing process as compared to the fermionic $4\rightarrow 2$ process, allows for a wider range of parameter space for producing the observed abundance while remaining within the self-interacting region. As depicted in Fig. \ref{vectorDMplot}, the scalar self coupling can take values roughly within the range $\lambda_{\rm s}\in(0.01,1)$, while the DM mass spans the range from tens of keV to a few MeV. The hidden sector gauge coupling takes values from few times $10^{-3}$ to $\sim 0.1$.

We note that in comparison to the fermion case, the viable values of the scalar self-coupling are significantly smaller and will remain perturbative under the renormalization group evolution up to the Planck scale.

\section{Conclusions and outlook}
\label{conclusions}

In this work we have studied how the dark matter relic density arises in scenarios where a scale invariant hidden sector interacts only feebly with the Standard Model degrees of freedom via a Higgs portal $\lambda \Phi^\dagger\Phi s^2$. We carried out thorough analysis of DM production in several benchmark scenarios where the hidden sector contains
either a scalar, fermion (sterile neutrino), or vector DM candidate. We found that requiring observable self interaction cross section and correct relic abundance essentially fixes the parameters of the scale invariant hidden sector, and results in an interplay between the electroweak scale, DM mass and couplings, reminiscent of the celebrated WIMP miracle.

The FIMP scenario remains not only as a viable but increasingly appealing mechanism for explaining the DM abundance since in this scenario the DM candidate is expected to leave no signatures in collider or direct detection experiments\footnote{This hindrance can be circumvented if cosmological history is altered from the usual radiation dominated case during the DM production and larger values of $\lambda_{\rm hs}$ become allowed \cite{Co:2015pka, Evans:2016zau}.}. However, despite the feeble coupling between the two sectors, there are several possibilities for observing DM indirectly.
One possibility for observing hidden sector DM indirectly would be to allow for a non-vanishing mixing angle between the singlet fermion and the SM neutrinos \cite{Heikinheimo:2016yds}.

Another important consequence of sizeable hidden sector interactions and the resulting dark freeze-out is the modification of DM momentum distribution function from the original freeze-in case. Even though the original FIMP miracle calculation would in some cases give the final DM abundance about right anyway, the eventual thermalization of DM within the hidden sector may have a large effect on formation of structures at large scales. Dark matter momentum distribution has been studied in the context of frozen-in sterile neutrinos in \cite{Merle:2015oja, Konig:2016dzg}, and it would be interesting to see what effect would also a spin-1 vector boson with $m_{\rm A}\lesssim{\mathcal O}(1)$ MeV have on structure formation.

As the determination of DM momentum distribution function is crucial for solving the exact effect DM has on cosmic structure formation, including the effects we have discussed in this work in a consistent way may be important for determining the origin and properties of DM. This is especially important as this might be the only way to test models where DM interacts only feebly with the SM sector. Therefore, we plan to address these aspects in more detail in forthcoming publications.

\section*{Acknowledgements}
This work has been supported by the Academy of Finland, grant\# 267842. TT is supported by the U.K. Science and Technology Facilities Council grant ST/J001546/1.

\bibliography{FIMP_miracle.bib}

\end{document}